# Gate-tunable quantum acoustoelectric transport in graphene

Yicheng Mou[1], Haonan Chen[1], Jiaqi Liu[1], Qing Lan[1], Jiayu Wang[1], Chuanxin Zhang[2], Yuxiang Wang[1], Jiaming Gu[1], Tuoyu Zhao[1], Xue Jiang[2*], Wu Shi[1,3*] and Cheng Zhang[1,3*]

[1] State Key Laboratory of Surface Physics and Institute for Nanoelectronic Devices and Quantum Computing, Fudan University, Shanghai 200433, China
[2] Center for Biomedical Engineering, School of Information Science and Technology, Fudan University, Shanghai 200433, China
[3] Zhangjiang Fudan International Innovation Center, Fudan University, Shanghai 201210, China

[*] Correspondence and requests for materials should be addressed to X. J. (E-mail: xuejiang@fudan.edu.cn), W. S. (E-mail: shiwu@fudan.edu.cn) & C. Z. (E-mail: zhangcheng@fudan.edu.cn).

**Abstract**

**Transport probes the motion of quasiparticles in response to external excitations. Apart from the well-known electric and thermoelectric transport, acoustoelectric transport induced by traveling acoustic waves has been rarely explored. Here, by adopting a hybrid nanodevices integrated with piezoelectric substrates, we establish a simple design of acoustoelectric transport with gate tunability. We fabricate dual-gated acoustoelectric devices based on BN-encapsuled graphene on $LiNbO_3$. Longitudinal and transverse acoustoelectric voltages are generated by launching pulsed surface acoustic wave. The gate dependence of zero-field longitudinal acoustoelectric signal presents strikingly similar profiles as that of Hall resistivity, providing a valid approach for extracting carrier density without magnetic field. In magnetic fields, acoustoelectric quantum oscillations appear due to Landau quantization, which are more robust and pronounced than Shubnikov-de Haas oscillations. Our work demonstrates a feasible acoustoelectric setup with gate tunability, which can be extended to the broad scope of various Van der Waals materials.**

In metals and semiconductors, directional motion of electrons can be generated by applying an electric bias or a temperature gradient. The acceleration of electron is then retarded by the scattering with impurities or phonons, resulting in an equilibrium state with finite electric potential drop in the system. Such transport process provides a simple yet versatile approach for probing electronic states, and has been established as one of the most important experimental techniques since the early stage of condensed matter physics[1,2]. The introduction of magnetic fields in transport further generates magnetoresistance and Hall effect, which offers a precise way of evaluating carrier concentration and sensing magnetic field as commonly used in semiconductor industry[1]. Moreover, in high-mobility systems, Shubnikov-de Haas (SdH) oscillations can be detected in resistivity at low temperatures due to the formation of discrete Landau levels[3]. It allows a direct measurement of Fermi surface properties such as Fermi wave vector, quasiparticle effective mass, and quantum lifetime. The investigation of quantum oscillations has played a vital role in the research of two-dimensional electron systems and topological materials in recent years[4–9].

For transport experiments, different types of excitations lead to distinct responses of quasiparticle motion. The electric bias only impacts charged carriers and results in a shift of the whole Fermi surface with momentum displacement. In contrast, the temperature gradient causes an entropy flow with the directional thermomigration of both charged and charge-neutral quasiparticles. The ratio of electric conductivity to the electronic contribution of thermal conductivity can be used to diagnose the non-Fermi liquid property beyond the Wiedemann-Franz law[10,11]. Therefore, the combination of multiple transport methods, including electric, thermal, and thermoelectric conductivity, could help explore different aspects of material properties in a more diverse way. It has been proven to be particularly advantageous helpful in many crucial cases such as pseudo-gap states of high-$T_c$ superconductors[12–14], quantum critical points[15], electron hydrodynamics[16,17], as well as novel charge-neutral quasiparticles[18,19].

Apart from conventional transport experiments with electric or thermal excitations, new types of transport technique have also been pursued over the years. Among them, a mechanical version of carrier excitation method, known as acoustoelectric (AE) transport, was proposed back in the 1950s[20]. It employs a traveling acoustic wave to generate a net momentum in the carriers, and measures the longitudinal as well as transverse voltages. To help distinguish these three types of carrier transport methods, Figure 1a-f illustrate the differences among electric, thermoelectric and AE transport phenomena, which leads to distinct impacts on motion of electrons and holes in magnetic fields as well as particle distributions in momentum space[1,21]. With the application of electric fields, electrons and holes drift towards opposite directions longitudinally, while their transverse motion in magnetic fields is the same (Fig. 1a). For temperature gradient, both carriers move together towards the same direction from $T_h$ to $T_l$, while their transverse motion in magnetic fields is opposite (Fig. 1b). For AE transport, the propagation of acoustic wave generates periodical band modulation to the conductive channel spatially. Both carriers are trapped in the energy extrema position (the minima of the conduction band for electrons, and the maxima of the valence band for holes). Then the carriers will be collectively dragged by the acoustic wave along the same direction due to the momentum transfer from phonons to electrons/holes. Similar to thermoelectric transport, the transverse motion of AE transport is opposite for electrons and holes (Fig. 1c). Figures 1d-f illustrate the change of finite-temperature Fermi-Dirac distribution and the corresponding Fermi surface in the presence of different excitations. The electric bias induces a net momentum shift while the temperature gradient excites electrons below the Fermi energy to higher energy with directional diffusive motion. As for acoustic excitation, the main contribution is from the interaction between the accompanied piezoelectric fields and electrons, which induces a high-frequency oscillation to the Fermi surface. Owing to the traveling nature of the applied acoustic wave, a net momentum shift will be generated to the Fermi surface. Note that the net momentum shift from the traveling acoustic wave differs from the direct momentum shift due to electric bias in two aspects: (1) the former is a second-order effect from the Fermi surface oscillation in piezoelectric fields; (2) the former has the same sign for electrons and holes while the latter is opposite.

Typically, surface acoustic wave (SAW) in the sub-gigahertz frequency is preferred due to its high intensity and low loss at propagation. Owing to the difficulty in integrating materials with piezoelectric substrates, AE transport had been limited to very few thin film materials[22,23] which can be epitaxially grown on certain wafers like GaAs or $LiNbO_3$, for a long time. The development of transfer technique in recent years provides the opportunity to fabricate nanodevices of Van der Waals materials on top of various substrates. AE transport has been carried out in graphene[24,25], $MoS_2$[26], and black phosphorus[27], indicating its potential in the research of Van der Waals materials. More interestingly, acoustic wave could modulate the electronic states and induce many exotic phenomena such as zero-field transverse voltage[28] and negative resistance state[29]. However, the combination of AE transport with other tuning knobs such as magnetic field and gate voltage remains to be explored. In this work, we adopt a handy design of dual-gated AE device based on hexagonal-boron-nitride (hBN)-encapsuled graphene on $LiNbO_3$, and systematically investigate the longitudinal and transverse components of AE response in magnetic fields.

Figure 2a shows the schematic of graphene AE device used in this study. Figure 2b and 2c are the cross-sectional view of device configuration and the optical image of a typical device, respectively. Firstly, the delay-line-type interdigital transducers (IDTs) were fabricated on commercially available black $LiNbO_3$ substrates (Y-cut or 128°-Y-cut) by photolithography and metal sputtering, which were designed to generate SAW propagating along the crystal's $z$-axis. Then we performed the dry-transfer method using polycarbonate (PC)/polydimethylsiloxane (PDMS) to transfer stacked BN/graphene/BN samples on $LiNbO_3$ in between IDTs[30,31]. Then the samples were patterned into Hall-bar and edge-contacted with electrodes (5nm Cr and 80nm Au). The electrode pattern was carefully designed to avoid large reflection of SAW. The top BN and bottom $LiNbO_3$ substrate serve as the top and back gate dielectrics, respectively. The thickness of the graphene device was verified using Raman spectrum as shown in Supplemental Material Fig. S1.

Magnetotransport measurements were performed in commercial variable temperature inserts with superconducting magnet (1.6-300 K, ±12 T). The IDTs were firstly characterized by a vector network analyzer. We used lock-in amplifiers to measure four-terminal longitudinal voltages and transverse voltages in both electric and AE transport experiments to avoid the influence of contact resistance. The "four-terminal AE voltage" refers to AE voltages measured from the four-terminal voltage-sensing electrodes of a Hall-bar device. In electric transport, 10 nA current with a low frequency of 3-17.777 Hz was applied between the source and drain electrodes. In AE transport measurements, a signal generator was used to output a pulsed-amplitude-modulated (PAM) microwave with a sub-gigahertz frequency $f_{SAW}$ to excite IDTs. The pulse width was 100 ns and the pulse period was 1 μs (which corresponds to a duty cycle of 10%) with a low modulation frequency of 3-17.777 Hz. The waveforms of PAM microwave were visualized by oscilloscope (a typical waveform example is shown in Supplemental Material Fig. S2). In this way, a series of pulsed SAW with amplitude modulation were generated.

Figure 2d presents the reflection spectrum ($S_{11}$) and transmission spectrum ($S_{21}$) in logarithmic scale at different temperatures, in which the center frequency could be determined. The frequency dependence of $S$-parameters is nearly unchanged in the range of 2-60 K while the corresponding curve of 300 K shows a distinct frequency shift. The center frequency evolves from 287 MHz at 2 K to 282 MHz at 300 K, it results from the shift of sound velocity in $LiNbO_3$ as temperature varies. Figure 2e is the measured frequency-dependent longitudinal AE voltages ($V_{xx}^{\mathrm{AE}}$) from the graphene channel under the SAW excitation at zero gate voltage. Unless otherwise noted, the output microwave powers were fixed at -10 dBm (0.1 mW). These curves present pronounced peaks at the center frequency of IDTs, at which the longitudinal AE voltage could reach several tens of microvolts. The AE signals switch sign as the SAW direction is reversed, consistent with the acousto-drag nature of carrier transport.

We further investigate the gate-dependence of AE voltages. Figure 3a presents transfer curves of graphene channel resistance by sweeping top-gate voltage and back-gate voltage (the inset of Fig. 3a) in Device 1, respectively. It shows clear ambipolar behavior with a sharp resistance peak over 2 kΩ at 2 K. The top panel of Fig. 3b is the gate dependence of $V_{xx}^{\mathrm{AE}}$ at zero magnetic field and 2 K. Here L→R and R→L denote the SAW propagating direction from left to right and right to left, respectively. For a fixed SAW propagating direction such as L→R (blue curve in Fig. 3b), $V_{xx}^{\mathrm{AE}}$ show dispersive-resonance-like features near charge neutral point with the absolute value rapidly decreasing towards high electron/hole doping. By switching the SAW direction, the curve of $V_{xx}^{\mathrm{AE}} - V_{\mathrm{TG}}$ present identical behavior but with opposite sign. It also resembles the gate dependence of thermopower in graphene[32]. For comparison, we present the gate dependence of Hall resistance $R_{yx}$ at ±1 T in Fig. 3b, which shows similar profile. It reveals an intimate connection between AE voltages and Hall resistance. Figure 3c presents the $f_{\mathrm{SAW}}$ - $V_{\mathrm{TG}}$ mapping diagram of $V_{xx}^{\mathrm{AE}}$. The two hot spots of $V_{xx}^{\mathrm{AE}}$ (yellow and blue regions) in the mapping diagram match well with the large peak/dip features close to charge neutrality in Fig. 3b. The rapid decrease of $|V_{xx}^{\mathrm{AE}}|$ away from the center frequency further verifies the AE mechanism. As shown in Fig. 3d, $V_{xx}^{\mathrm{AE}}$ presents a linear dependence on the SAW excitation power at different gate voltages.

To quantitatively analyze the AE effect, we present a semiclassical formula for the longitudinal AE current density[22,25,28] as below:

$$j_{xx}^{\mathrm{AE}} = \frac{\mu I \Gamma}{v} \tag{1}$$

where $\mu$ is carrier mobility, $I$ is the intensity of acoustic wave, $\Gamma$ is the attenuation coefficient, and $v$ is the acoustic velocity, for Y-cut Z-propagate $LiNbO_3$ $v$ = 3488 m/s[33,34]. For conductors intimately lying on the piezoelectric substrate, the attenuation coefficient for surface acoustic wave could be expressed as:

$$\Gamma = \frac{\pi K^2}{\lambda} \frac{\sigma_{xx}/\sigma_m}{1+(\sigma_{xx}/\sigma_m)^2} \tag{2}$$

where $K^2$ refers to the electromechanical coupling coefficient of the piezoelectric substrate ($K^2$ = 4.0% for Y-cut Z-propagate $LiNbO_3$[33,34]), $\lambda$ is the wavelength of SAW, and $\sigma_{xx}(\omega)$ is the frequency-dependent sheet conductivity[35,36]. And $\sigma_m = v\varepsilon_0(1+\varepsilon_r)$ is a substrate-dependent characteristic sheet conductivity, with $\varepsilon_0$ and $\varepsilon_r$ being permittivity of vacuum and relative dielectric coefficient of the substrate, respectively ($\sigma_m = 1.39 \times 10^{-6}$ S for Y-cut $LiNbO_3$). Since here the graphene sheet conductivity is in the order of $10^{-3}$ S for the whole gate range ($\sigma_{xx} \gg \sigma_m$), the equation of AE current density can then be reduced to:

$$j_{xx}^{\mathrm{AE}} = \frac{\pi K^2 I \sigma_m}{\lambda v} \frac{1}{ne} \quad (3)$$

It nicely accounts for the strong similarity between Hall resistance ($R_{yx} \propto \frac{1}{ne}$) and AE voltages observed in Fig. 3b, as well as the power dependence of $V_{xx}^{\mathrm{AE}}$ in Fig. 3d. Equation (3) further allows the extraction of carrier density from the AE transport result. Figure 3e presents the extracted carrier density from AE transport and Hall effect, which generally agrees with each other.

We also performed dual-gate sweeping as shown in Fig. 3f. It allows an independent control of carrier density $n$ and displacement electric field $D$, where $n = (C_{\mathrm{TG}}V_{\mathrm{TG}} + C_{\mathrm{BG}}V_{\mathrm{BG}})/e$, and $D = (-C_{\mathrm{TG}}V_{\mathrm{TG}} + C_{\mathrm{BG}}V_{\mathrm{BG}})/2$. Here $C_{\mathrm{TG}}$ and $C_{\mathrm{BG}}$ are the capacitance per unit area of the top- and back-gate layers[37,38]. The resistance peak in the scan of $V_{\mathrm{TG}}$ systematically shifts as $V_{\mathrm{BG}}$ varies. Similarly, the $V_{\mathrm{TG}}$ position, where the sign change of $V_{xx}^{\mathrm{AE}}$ occurs, also evolves with $V_{\mathrm{BG}}$. In the meantime, the maximum absolute value of $V_{xx}^{\mathrm{AE}}$ shown in the bottom panel of Fig. 3f decreases towards large $V_{\mathrm{BG}}$, consistent with the decrease of resistance peak in the top panel. It indicates that the presence of a displacement field leads to a decrease of resistance peak at charge neutral point. Generally, it can be considered as the result of mobility enhancement due to the screening of Coulomb impurities as well as suppression of charge puddles at large displacement field. Consequently, the peak of $V_{xx}^{\mathrm{AE}}$ also systematically decreases with the increase of displacement field for a fixed carrier density since $V_{xx}^{\mathrm{AE}} \propto j_{xx}^{\mathrm{AE}} R_{xx}$. The mobility (or conductivity) enhancement of graphene leads to a change in the dielectric environment, which in turn affects the local electromechanical coupling coefficient as well as acoustic velocity. It can induce the nonlinear dependence of acoustic attenuation on the channel resistance. Meanwhile, multi-carrier transport may also happen at large doping regions for graphene. Therefore, the carrier density calculated from the acoustoelectric effect and the Hall effect slightly deviate at large gate voltages, as observed in Fig. 3e.

In the presence of external magnetic fields, electrons in graphene would generate transverse motion and form cyclotron orbits, which leads to discrete Landau levels in the energy bands[6,39]. We summarize the comparison of AE transport and electric transport in Fig. 4. Figure 4a-b are the magnetic field dependence of longitudinal and transverse AE voltages ($V_{xx}^{\mathrm{AE}}$ and $V_{yx}^{\mathrm{AE}}$) at different temperatures, while Figs. 4c-d are that of longitudinal resistance and Hall resistance ($R_{xx}$ and $R_{yx}$). Similar to electrical and thermoelectric transport, the longitudinal AE voltage is even in magnetic fields, while the transverse one is odd. Strong quantum oscillations appear in both AE and electric transport signals as a result of Landau quantization. The amplitude of oscillations systematically decreases as the temperature rises. The high oscillation frequency indicates a large Fermi surface in the graphene sample. We note that the data in Figs. 4-5 was collected from Device 2 on a 128°-Y-cut $LiNbO_3$ substrate. The large Fermi surface results from the charge doping effect due to the out-of-plane component of ferroelectric polarization in 128°-Y-cut $LiNbO_3$. We intentionally choose to the highly-doped sample to highlight the difference in quantum oscillations between AE and electric transport. Quantum oscillations in AE transport persists above 60 K, while SdH oscillations in electric transport becomes negligible at lower temperatures. Meanwhile, the oscillation amplitude in AE transport is much stronger. At 2 K, the on-set field for AE quantum oscillation is about 1.4 T, while SdH oscillations start around 3 T. It indicates that AE transport is more sensitive to Landau levels than electric transport.

We further investigate quantum oscillations in AE voltages in details as shown in Fig. 5. Figure 5a is an enlarged view of $V_{xx}^{\mathrm{AE}}$ and $V_{yx}^{\mathrm{AE}}$ at 2 K. The quantum oscillation of AE voltages can also be detected by sweeping the gate voltage as shown in Fig. 5b. In Fig. 5c, we present the fast Fourier transform (FFT) analysis results on the extracted AE oscillations (in the region of 4.2-8.3 T) at different temperatures. The inset of Fig. 5c is the fit of FFT amplitude damping with temperature following the thermal damping factor $R_T = \frac{2\pi^2(k_B T/\hbar\omega)}{\sinh(2\pi^2 k_B T/\hbar\omega)}$ of Lifshitz-Kosevich (LK) formula. Here the cyclotron frequency is $\omega = eB/m^*$ with $m^*$ being the effective mass. It yields an effective mass value of $0.049m_e$, where $m_e$ is the rest mass of electron. We note that unlike the microwave-induced quantum oscillations observed in two-dimensional electron gases[40], the current SAW frequency is well below the resonant value between Landau levels in graphene (in the order of $10^{12}$ Hz). System with much larger effective mass may be able to achieve resonant-field-induced quantum oscillations in the frequency range of SAW. Figure 5d is the Landau fan diagram fitting using the peak and valley positions in AE oscillations. It gives an oscillation frequency of 58.5 T, consistent with the dominant peak in the FFT results. In Fig. 5e, we present a comparison of extracted quantum oscillations in $V_{xx}^{\mathrm{AE}}$ $R_{xx}$, and $V_{yx}^{\mathrm{AE}}$ by subtracting their background in magnetic fields. Surprisingly, we find that the peaks and valleys of $\Delta R_{xx}$ is close to zero points of $\Delta V_{xx}^{\mathrm{AE}}$, which points to a $\pi/2$ phase shift between SdH oscillations and AE oscillations. It is likely to originate from the fact that AE transport is a second-order effect from the Fermi surface oscillation in alternating electric fields of SAW, while electric transport is a first-order consequence of the Fermi surface shift in electric fields (see the comparison provided in Fig. 1). Such behavior also matches a previous theoretical prediction based on the phenomenological model of the acoustoelectric drag effect[41]. The model also explains why AE transport is more sensitive to Landau levels as observed in Fig. 4. Based on this model, the AE current density $j_\alpha$ ($\alpha = x$ or $y$, corresponding to longitudinal and transverse component, respectively) in the presence of magnetic field is proportional to $\partial\sigma_{\alpha x}/\partial\nu$, where $\sigma_{\alpha x}$ is the components of conductivity tensor and $\nu$ is the filling factor. It indicates that the magnitude of AE transport is proportional to the first derivative of conductivity, which helps to subtract the field-dependent magnetoresistance background (usually very large for high-mobility systems) and therefore enlarge the oscillating component. It also gives the $\pi/2$ phase shift between SdH oscillations and AE oscillations. And the phase shift between $V_{xx}^{\mathrm{AE}}$ and $V_{yx}^{\mathrm{AE}}$ inherits from that of $R_{xx}$ and $R_{yx}$, which varies with the ratio of $R_{xx}/R_{yx}$ as well as detailed material properties[42,43]. These properties are very similar to the case of thermoelectric effect, which is also related to the derivative of conductivity through the Mott relation and considered as a more sensitive probe for Landau levels. In addition to quantum oscillations, there were also discussions over whether transverse AE current will be quantized at strong magnetic fields for two-dimensional electron systems[41,44]. The observation of AE quantum oscillations in graphene offers a test for these theories.

The gate-tunable AE transport presented in this work has unveiled its potential in probing electronic states of condensed matters. Compared with early studies of AE transport using bulk acoustic wave, AE transport technique based on SAW on $LiNbO_3$ substrates enables an efficient generation of highly directional and low-dissipation acoustic waves. Moreover, $LiNbO_3$ substrates have a much larger electromechanical coupling coefficient than GaAs previously used in AE transport study of two-dimensional electron gases. Other than only working as an experimental probe, recent studies have also reported the manipulation of excitons[31,45,46], magnons[47], as well as skyrmions[48,49] through electron-phonon or magnetoelastic coupling with SAW. Such electronic modulation mechanism may be extended to AE transport as well. Recently, Y. W. Fang *et al.* also employed SAW resonators with high $Q$ value to enhance the attenuation signal, and measured the velocity shift using high-frequency lock-in technique[50]. It allows the extension of acoustic attenuation previously used in thin films of two-dimensional electron gases[35,51,52] to nanoscale materials such as exfoliated graphene. Together with our work on the AE effect, these progresses demonstrates that the application of SAW can serve as a versatile tool for condensed matter research.

In summary, we carried out the AE transport in hBN-encapsulated graphene on $LiNbO_3$ at gate voltages and magnetic fields. By investigating the gate-dependent longitudinal AE voltages, we reveal the intimate connection between AE voltages and Hall resistance. A valid approach for extracting carrier density from AE transport without the need of external magnetic field is demonstrated. It may be useful for intrinsic semiconductors or strong ferromagnets, in which Hall effect is hard to be detected due to large longitudinal resistivity or anomalous Hall resistivity. In the presence of external magnetic fields, quantum oscillations can be detected in longitudinal and transverse AE voltages. With the combination of gate and magnetic field modulations, AE transport could serve as an important complement for electric and thermoelectric transport. The easy integration of piezoelectric substrates through the transfer technique greatly enhances the accessibility of AE transport in the research of Van der Waals materials.

**Supporting Information**

Sample characterization by Raman spectrum, visualization of pulsed amplitude-modulated surface acoustic waves, comparison of open-circuit and short-circuit configuration, temperature dependent $S$-parameters of IDTs, derivation from equation (1) to (3).

**Author contributions**

C.Z. conceived the idea and supervised the overall research. Y.M., H.C. and J.L. fabricated the devices assisted by Q.L., J.W., C.X.Z., and T.Z. Y.M. carried out the electrical transport measurements with the help of Y.W. and J.G. Y.M. and J.L. performed Raman spectrum measurement. Y.M., C.Z., W.S. and X.J. analyzed the data. Y.M. and C.Z. prepared the manuscript with contributions from all other co-authors.

**Notes**

The authors declare no competing financial interests.

**Acknowledgements**

The authors would like to thank Wenbin Wang, Hua Jiang, Dongsheng Liu, Zhipeng Zhong, Yang Shi and Jingru Li for their beneficial discussions and constructive suggestions. C.Z. was sponsored by the National Key R&D Program of China (Grant No. 2022YFA1405700), the National Natural Science Foundation of China (Grant No. 92365104 and 12174069), and Shuguang Program from the Shanghai Education Development Foundation. W.S. was supported by the National Natural Science Foundation of China (Grant No. 12274090) and the Natural Science Foundation of Shanghai (Grant No. 22ZR1406300). X.J. was sponsored by the National Key R&D Program of China (Grant No. 2023YFA1407800), and the National Natural Science Foundation of China (Grant No. T2222024). Part of the sample fabrication was performed at Fudan Nano-fabrication Laboratory.

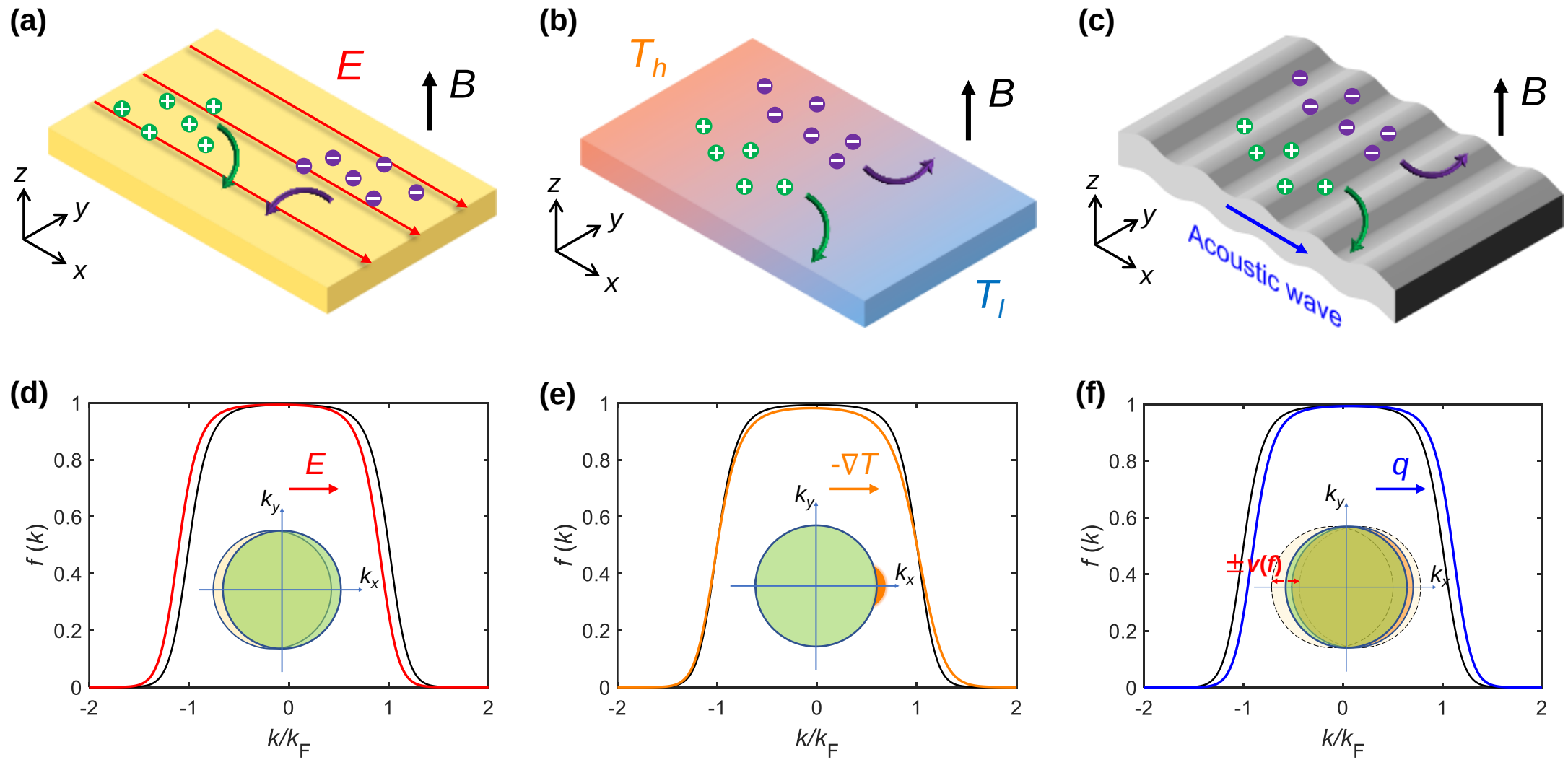

**Figure 1.** Illustration of the physical process of electric, thermoelectric and acoustoelectric transport. (a) Motion of electrons (purple balls) and holes (green balls) driven by electric field **E** with the existence of an external magnetic field **B**. (b) Motion of electrons and holes driven by a temperature gradient -∇$T$ in magnetic fields. (c) Motion of electrons and holes driven by an acoustic wave in magnetic fields. (d) The finite-temperature Fermi-Dirac distribution with (red curve) and without (black curve) electric field. The insert figure presents a shift of the Fermi surface under **E**. (e) The finite-temperature Fermi-Dirac distribution with (orange curve) and without (black curve) temperature gradient. The insert figure shows the net momentum generated by -∇$T$. (f) The finite-temperature Fermi-Dirac distribution with (blue curve) and without (black curve) acoustic wave. The insert figure demonstrates that acoustic waves make Fermi surface oscillate periodically, and the mechanism of momentum deliver from acoustic wave to carriers will results a net shift of Fermi surface.

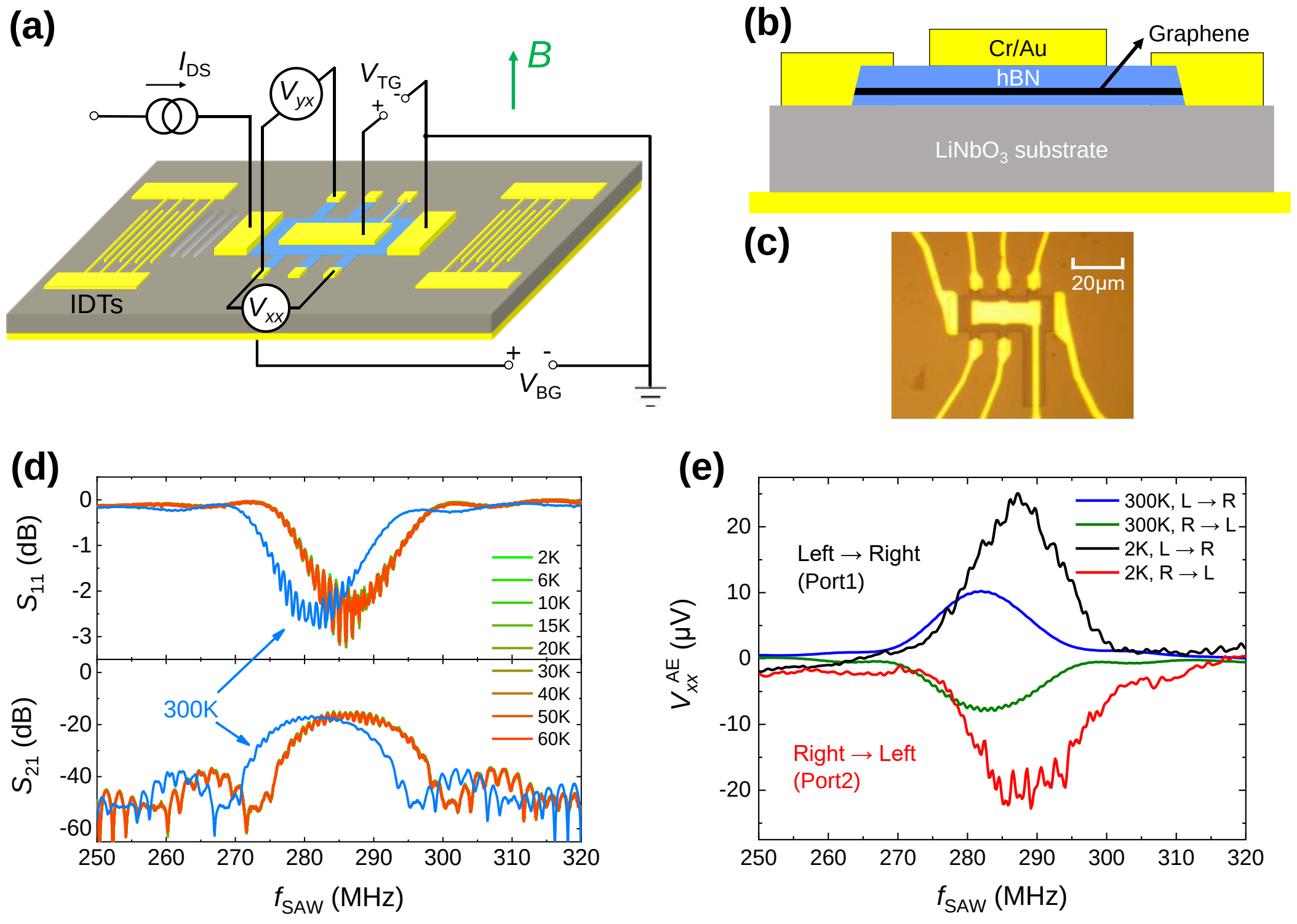

**Figure 2.** Device schematic and frequency dependence of device characterizations. (a) Schematic representation of our hybrid device with Hall-bar geometry. The blue region represents the hBN/graphene/hBN heterostructure, and the yellow region represent Cr/Au electrodes. The Van der Waals heterostructure is placed in the center of the delay line and connected to top and back gate electrodes. (b) Cross-sectional view of our device. (c) Optical image of one of the hBN-encapsulated devices, the distance between two adjacent electrodes is about 10 μm. (d) *S*-parameters of delay-line-type IDTs device on frequency domain. (e) Frequency dependence of longitudinal acoustoelectric voltage at 300 K and 2 K under distinct direction of surface acoustic waves, the input RF power is fixed at -10 dBm (0.1 mW). Here "port" means the microwave input port. Port1 corresponds to surface acoustic waves from left to right, and Port2 corresponds to the reverse surface acoustic waves from right to left.

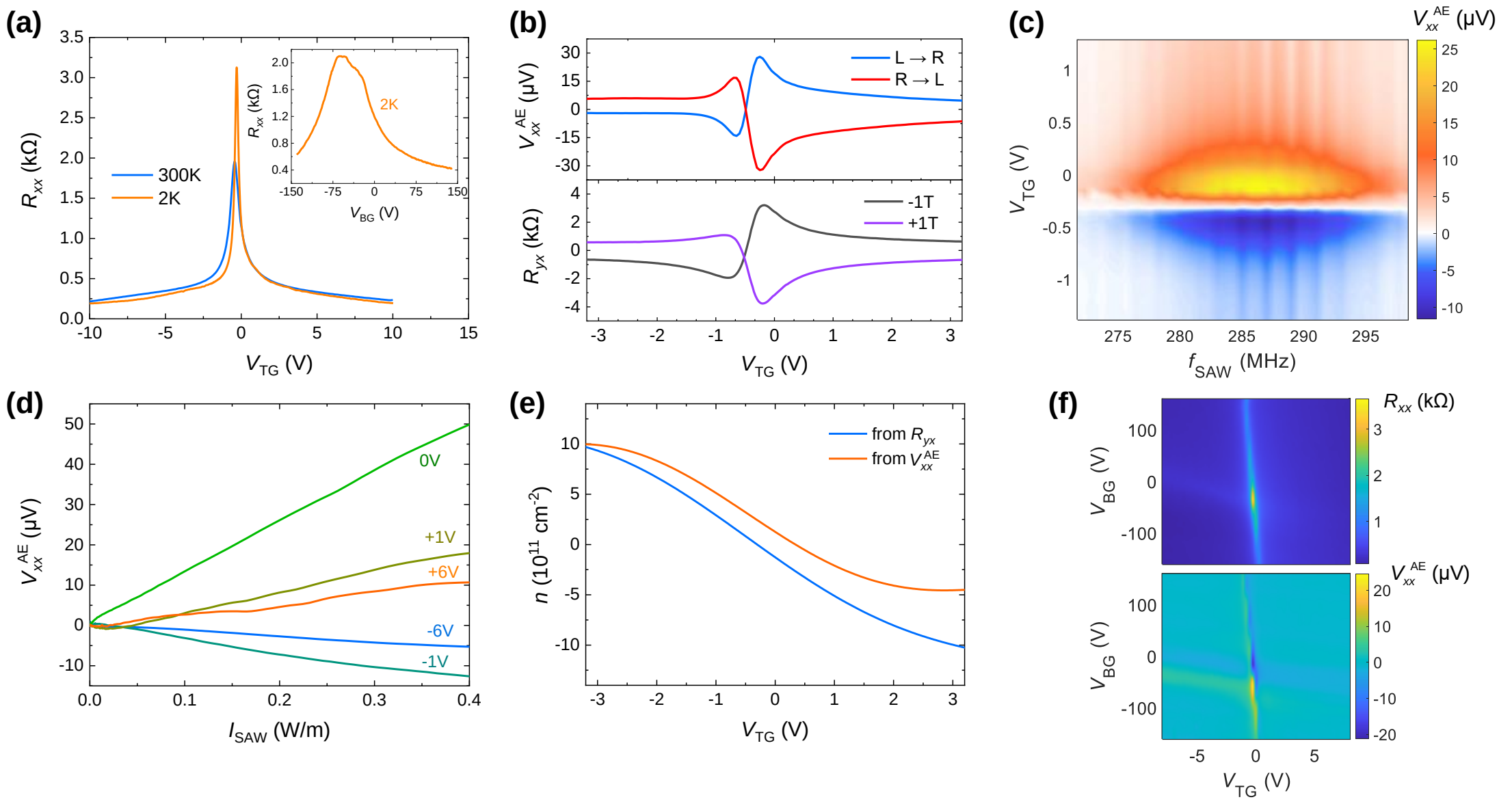

**Figure 3.** Results of gate-tunable acoustoelectric effect. (a) Transfer curves of graphene versus top gate voltage at 300 K and 2 K. Inset: transfer curves of graphene versus back gate voltage at 2 K. (b) Gate voltage dependence of longitudinal AE voltages $V_{xx}^{\mathrm{AE}}$ and Hall resistance $R_{yx}$ at ±1 T, the temperature is at 2 K, the acoustic frequency is 287 MHz and the input RF power is fixed at -10 dBm. The AE voltages and Hall resistance both show butterfly-shaped curves and become to zero consistently. (c) $V_{xx}^{\mathrm{AE}}$ as a function of frequency and top gate voltage at 2 K. The maximum and minimum values appear on both sides of the charge neutral point at center frequency. And at charge neutral point, the AE voltage goes to zero within the whole frequency range. (d) $V_{xx}^{\mathrm{AE}}$ as a function of surface acoustic wave intensity under different gate voltages at 2 K. (e) Carrier density calculated from $V_{xx}^{\mathrm{AE}}$ and $R_{yx}$, respectively. (f) Four terminal longitudinal resistance $R_{xx}$ and longitudinal AE voltage $V_{xx}^{\mathrm{AE}}$ as a function of top gate and back gate voltages at 2 K.

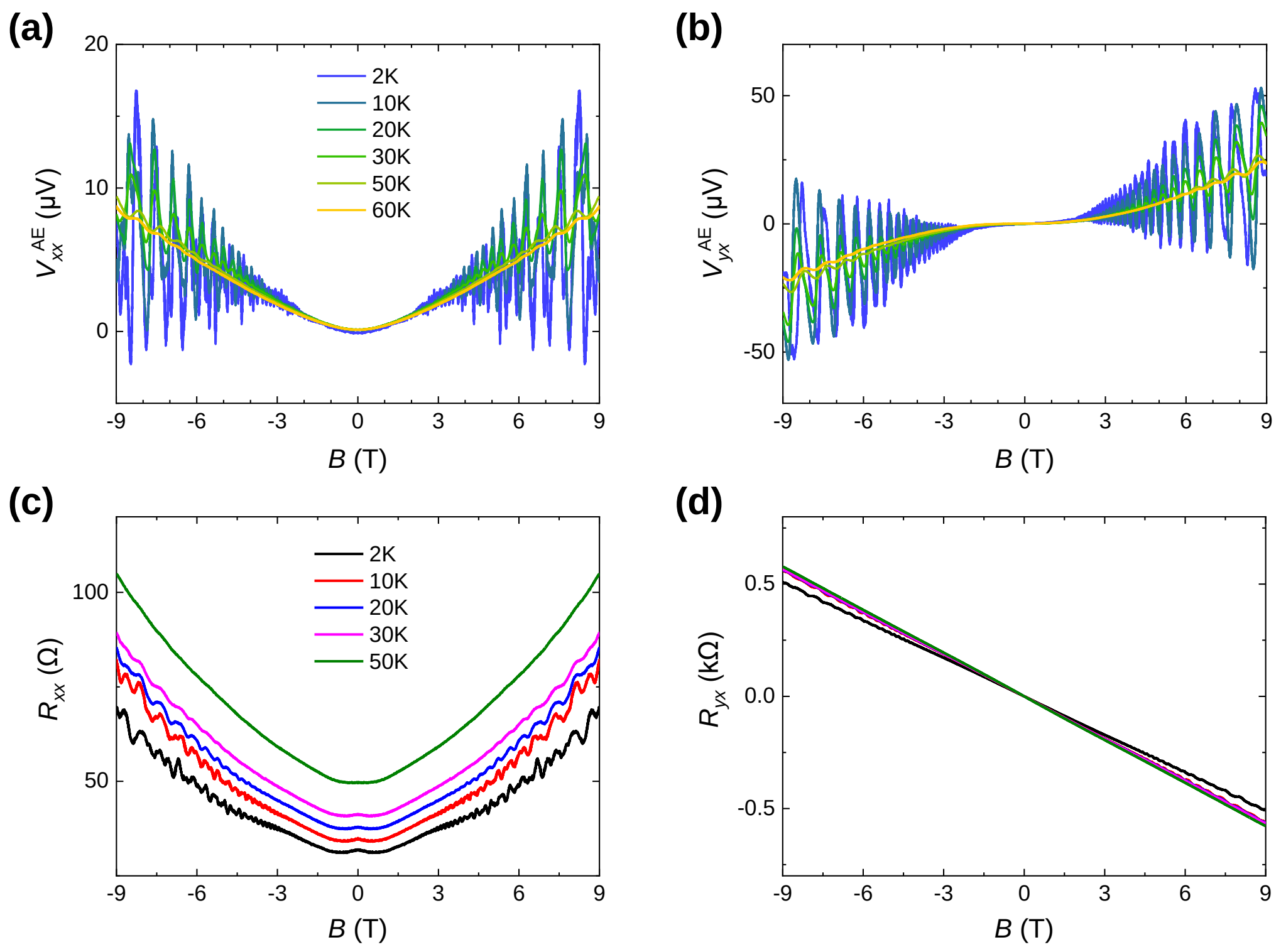

**Figure 4.** Magnetic field dependence of longitudinal and transverse response in AE and electric transport. (a) Four-terminal longitudinal AE voltages $V_{xx}^{\mathrm{AE}}$ versus magnetic field at 2-60 K. (b) Transverse AE voltages $V_{yx}^{\mathrm{AE}}$ versus magnetic field at 2-60 K. (c) Four-terminal longitudinal magnetoresistances $R_{xx}$ at 2-50 K. (d) Hall resistances $R_{yx}$ at 2-50 K.

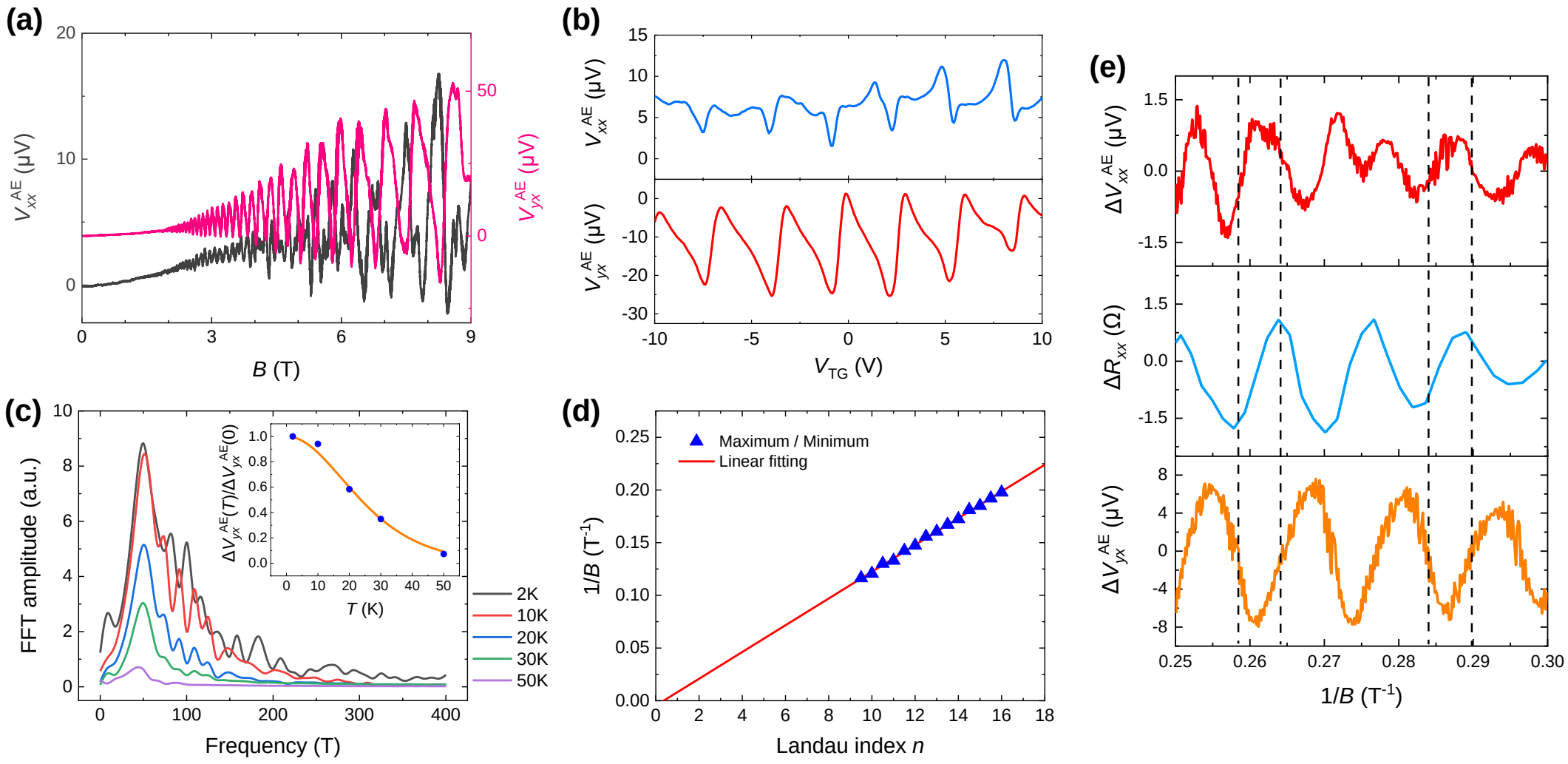

**Figure 5.** Quantum oscillations on AE voltages in magnetic fields. (a) Four-terminal longitudinal AE voltage $V_{xx}^{\mathrm{AE}}$ and transverse AE voltage $V_{yx}^{\mathrm{AE}}$ as a function of magnetic field at $T = 2$ K. The frequency is fixed at 313 MHz and RF power is fixed at -10dBm. (b) $V_{xx}^{\mathrm{AE}}$ and $V_{yx}^{\mathrm{AE}}$ as a function of top gate voltage at 9 T at 2 K. The acoustoelectric voltages oscillate as the top gate voltage changes due to Fermi energy passing from one Landau level to the other. (c) FFT spectrum of the quantum oscillations from $V_{yx}^{\mathrm{AE}}$ after subtracting the background. The inset is the LK fitting based on the FFT amplitude. (d) Landau fan diagram of quantum oscillation on AE voltages. (e) Stacked figures of $\Delta V_{xx}^{\mathrm{AE}}$, $\Delta R_{xx}$ and $\Delta V_{yx}^{\mathrm{AE}}$ versus $1/B$. The quantum oscillations of $\Delta V_{xx}^{\mathrm{AE}}$ and $\Delta R_{xx}$ have $\pi/2$ phase shift, and $\Delta V_{xx}^{\mathrm{AE}}$ and $\Delta V_{yx}^{\mathrm{AE}}$ have $\pi$ phase shift.